\documentclass[twocolumn,runningheads,natbib]{svjour2}
\smartqed  % flush right qed marks, e.g. at end of proof
\usepackage{graphicx}
\usepackage{mathptmx}      % use Times fonts if available on your TeX system
\newcommand{\be}{\begin{equation}}
\newcommand{\ee}{\end{equation}}
\newcommand{\bea}{\begin{eqnarray}}
\newcommand{\eea}{\end{eqnarray}}
\newcommand{\bann}{\begin{eqnarray*}}
\newcommand{\eann}{\end{eqnarray*}}
\newcommand{\bi}{\begin{itemize}}
\newcommand{\ei}{\end{itemize}}
\newcommand{\bcen}{\begin{center}}
\newcommand{\ecen}{\end{center}}

\journalname{Astrophysics and Space Science}
\begin{document}

\title{Spin-one color superconductivity in compact stars?- \\
an analysis within NJL-type models
\thanks{D.N.A work and attendance to the meeting was supported by VESF-Fellowships EGO-DIR-112/2005}
%{Grants or other notes
}


\author{D.N. Aguilera         
%\and        Second Author %etc.
}


\institute{D.N. Aguilera \at
              Department of Applied Physics \\
              Faculty of Sciences\\
              University of Alicante\\
              Tel.: +34-96-590-9599\\
              Fax: +34-96-590-9726\\
              \email{deborah.aguilera@ua.es}           %  \\
}

\date{Received: date / Accepted: date}

\maketitle

\begin{abstract}
We present results of a microscopic calculation using NJL-type model of possible spin-one pairings in two flavor quark matter for applications in compact star phenomenology. We focus on the color-spin locking phase (CSL) in which all quarks  pair in a symmetric way, in which color and spin rotations are locked. The CSL condensate is particularly interesting for compact star applications since it is flavor symmetric and could easily satisfy charge neutrality. Moreover, the fact that in this phase all quarks are gapped might help to suppress the direct Urca process, consistent with cooling models. The order of magnitude of these small gaps ($\simeq 1$ MeV) will not influence the EoS, but  their also small critical temperatures ($T_c \simeq 800$ keV) could be relevant in the late stages neutron star evolution, when the temperature falls below this value
and a CSL quark core could form.

\end{abstract}

\section{Introduction}
\label{intro}

The most favorable places in nature where color superconducting states  of 
matter are expected to occur are the interiors of compact stars, with 
temperatures well below 1 MeV and central densities exceeding the 
nuclear saturation density $\rho_0$ by several times.

Color superconductivity has been widely studied from non-perturbative low-energy QCD models 
where gaps of the order of magnitude of $\simeq 100$ MeV has been calculated (\cite{Rapp:1997zu,Alford:1997zt}). 
One of the effective models most used is the Nambu Jona-Lasinio (NJL) 
model that considers that the quarks  interact locally by a 4-point vertex effective force 
and  disregards the gluon degrees of freedom. 
The model uses an attractive interaction in the scalar 
%$\bar q q$ 
meson channel that 
causes spontaneous chiral symmetry breaking if the coupling is strong enough.  
At the mean field level, the model shows how quarks acquire a dynamical constituent mass, which is proportional to the vacuum expectation value of the scalar field. 
The NJL model has then been extended and widely used to described successfully chiral restoration at finite temperatures and color superconductivity at finite density (for a review see \cite{Buballa:2003qv}). 

For compact star applications, color super\-conduc\-ting quark matter phases enforcing color and charge neutrality 
has been widely studied (\cite{Alford:2000sx,Steiner:2002gx}). 
It has been shown that 
local charge neutrality may disfavor 
 the occurrence of phases with large gaps where quarks with 
different flavor pair in a spin-0 condensate, like the 2SC phase (\cite{Alford:2002kj}).  
On the other hand, NJL-type model calculations show that the intermediate density region of the neutral QCD phase diagram, where the quark chemical potential is not sufficiently large to have the strange quark deconfined ($\mu \geq 430-500$ MeV, see \cite{Buballa:2001gj,Neumann:2002jm}), might be dominated by $u,d$ quarks (\cite{Ruster:2005jc,Blaschke:2005uj}).  If this is the case, 2-flavor quark matter phases may occupy a large volume in the core of compact stars (\cite{Grigorian:2003vi,Shovkovy:2003ce}). 

While the stability of  neutral 2SC pure phase is rather model dependent and might be unlikely for moderate coupling constants (\cite{Aguilera:2004ag,GomezDumm:2005hy}),   
phases with other pairing patterns (or none at all) become important for the phenomenology of neutron stars. 
Then, besides other possibilities%
\footnote{no pairing (normal quark matter), pairing with displacement of the Fermi surfaces (LOFF or crystalline structure) or deformation of the Fermi surfaces , interior gap structure, gapless 2SC or gCFL, etc.}, quarks could pair in spin-one condensates (\cite{Schafer:2000tw,Alford:2002rz}). 

These condensates with small pairing gaps ($\Delta \simeq 1$ MeV), are not expected to have influence on the equation of state but could strongly affect the transport and thermal properties of quark matter and therefore leads to consequences for the phenomenology of compact stars.

An energy gap in the quasiparticle excitation spectrum introduces a suppression of the neutrino emissivity and the specific heat of the paired fermions by a Boltzmann factor $\exp(-\Delta/kT)$ for $T$ smaller than the critical temperature  $T_c$ for pair formation. 
Thus, since the neutrino luminosity is strongly dominated by the neutrino emission from the core, the cooling of a neutron star during its early life  will be affected by the dense matter pairing pattern. 
On the other hand, most of the specific heat of the star is provided by the core and it shows a discontinuity when the star temperature crosses $T_c$. A comparison of cooling curves with observations could provide a hint to discriminate dense matter phases.  Several studies have shown that the occurrence of unpaired quarks in the core leads to rapid cooling via the direct Urca process , incompatible with the observations. Thus, phases that present no gapless modes with small pairing gaps ($\Delta \simeq 1$ MeV) prevent the direct Urca to work uncontrolled and  might help to give a consistent picture of the observed data (see \cite{Page:2000wt,Page:2004,Yakovlev:2004,Grigorian:2004jq}).

On the other hand,  it has been demonstrated that the CSL phase exhibit a Meissner effect and the magnetic field of a neutron star would be  expelled from a CSL quark core if it does not exceed the critical magnetic field $B_{c}$ that destroy the superconducting phase \cite{Schmitt:2003xq}. This hypothesis might be consistent with recent investigations that indicate the crust confination of the magnetic field \cite{Geppert:2004sd,Perez-Azorin:2005ds}.

To study this qualitatively, we present in this work a summary of the results obtained in previous works 
of NJL model calculations 
of two spin-one pairing patterns, focusing in the CSL phase, and a new analysis of 
features that are relevant to apply the CSL in phenomenological applications, 
like compact star cooling or protoneutron star evolution.

\section{Brief on the model}
\label{sec:1}
We consider a two-flavor system of quarks, $q = (u,d)^T$, with an NJL-Lagrangian 
$\mathcal{L}_{{\rm eff}}= \mathcal{L}_0 + \mathcal{L}_{q\bar q}+\mathcal{L}_{qq}$. 
which contains a free part 
$\mathcal{L}_0=\bar q\left(i\not\!\partial-m \right)q~$, a quark-antiquark interaction channel  
$\mathcal{L}_{q\bar q}$ that corresponds to the condensate for each flavor $f$ 
\be
     \sigma_f = \langle \bar q_f q_f \rangle~,  
\label{sigmaf}
\ee
and a diquark channel $\mathcal{L}_{q q}$ with color superconducting condensate matrix $\hat \Delta$ (components $\Delta^i$) 
which we will made explicitly in Sec.~\ref{sec:2}. 
The condensates (\ref{sigmaf}) are responsible for dynamical chiral symmetry breaking in vacuum and 
define the constituent quark masses as 
\be
M_f=m_f-4G\sigma_f
\label{constituent_mass}
\ee
being $G$ the coupling constant in the meson scalar channel. 
In the density regime investigated, these condensates are
relatively small but their finite size have consequences in the dispersion relations (see discussion in {\it Absence of ungapped modes} in Sec.~\ref{sec_likely_2SC} and Fig~\ref{fig:3}). 

After performing a linearization of $\mathcal{L}_{{\rm eff}}$ in the presence of the condensates
and providing the inverse of the fermion propagator in Nambu-Gorkov space 
\be  
S^{-1}(p)= 
\left(  
\begin{array}{cc}  
 \not\!p +\mu_f\gamma^0-\hat M&  
\hat\Delta\\
-{\hat\Delta}^\dagger&
\not\!p -\mu_f\gamma^0-\hat M  
\end{array}  
\right)~,
\ee
 performing usual techniques of thermal field theory, the thermodynamical potential $\Omega(T,{\mu_f})$ can be derived. At the mean-field level, i.e. the
stationary points 
\be
    \frac{\delta\Omega}{\delta\Delta^i} = 0~, \quad
    \frac{\delta\Omega}{\delta M_f} = 0~,
\label{gap_equations}
\ee
define a set of gap equations for $\Delta^i$ and $M_f$. 
Among the solutions, the stable one is the solution
which corresponds to the absolute minimum of $\Omega$.

The parameters (current quark mass $m=m_u=m_d$,  the coupling $G$,  
 and three dimensional cut-off  
$\Lambda$) 
%are shown in Tab. \ref{parNJL}  and 
have been determined by fitting the pion mass  
and decay constant to their empirical values and to vacuum constituent 
quark mass at zero momentum, $M$. In this work we consider 2-flavor quark matter ($u,d$)  assuming that the $m_s$ is large enough to appear only at higher densities.

%Through the work we will refer also to results that correspond to 
In this work we will also reference results obtained in
nonlocal extensions of the NJL model. 
The idea basically is to modify the quark interactions in order to act over a certain range in the momentum 
space introducing momentum dependent form factors $g(p)$ in the 
the current-current interaction terms. We will skip the details on the model  but the main consequence 
is that the inclusion of high momenta states beyond the usual NJL-cutoff causes a reduction of 
the diquark condensates 
and a lowering of the chiral phase transition (for a complete discussion see \cite{Aguilera:2006cj}).

\section{Spin-one pairing for compact stars applications}
\label{sec:2}

\subsection{{\it Unlikely:} 2SCb phase} 

Besides the spin-0 isospin singlet condensate (2SC) of two colors 
(e.g. $red$ and $green$), the remaining unpaired color (say $blue$) could pair 
in an anisotropic spin-one channel  with an expectation value of 
$\zeta = \langle q^{T}~C\sigma^{03}\tau_2~\hat P ^{(b)}q \rangle$ (see \cite{Buballa:2002wy} for NJL and 
\cite{Aguilera:2005uf} for a nonlocal extension).
Then, in the 2SCb phase we assume
\bea
\hat\Delta^{2SCb}=\Delta(\gamma_5\tau_2\lambda_2)(\delta_{c,r}+\delta_{c,g})
+\Delta^{\prime}(\sigma^{03}~\tau_2~\hat P_3^{(c)})\delta_{c,b}~,
\eea
and the thermodynamical potential is
\bea
\Omega^{2SCb}(T,\mu)&=&
\frac{(M_f-m)^2}{4G}+\frac{|\Delta|^2}{4G_1}
+\frac{|\Delta^{\prime}|^2}{16G_2}\nonumber
\\
-
%V^{2SCb}-
4 \sum_{\pm,i=1}^{3} \int &&\frac{d^3p}{(2\pi)^3}
\left[
\frac{E_i^{\pm}}{2}
%+T \ln{(1+e^{-E_i^-/T})}
+T \ln{(1+e^{-E_i^{\pm}/T})}
\right]~.\nonumber\\
\label{Omegablue}
\eea
The dispersion relations for the 2SC-paired quarks ($r,g$)
\be
(E_{1,2}^{-})^2=(E^{-})^2=(\epsilon - \mu)^2 +|\Delta|^2~,
\ee
with the free particle energy $\epsilon^2= \vec p\,^2+M^2$ 
and for the third color ($b$) quarks
\be
(E_{3}^{-})^2 = (\epsilon_{\rm eff}- \mu_{\rm eff})^2 +|\Delta^{\prime}_{\rm eff}|^2
\ee 
where the effective variables depend  on the angle $\theta$, 
with $\cos \theta= p_3/|\vec p|$, and are defined as
\bea
\epsilon_{\rm eff}^2&=& \vec p\,^2+M_{\rm eff}^2~,\\
M_{\rm eff}&=&M\frac{\mu}{\mu_{\rm eff}}~,\\
\mu_{\rm eff}^2&=&\mu^2+|\Delta^{\prime}|^2\sin^2{\theta}~,\\
|\Delta^{\prime}_{\rm eff}|^2 &=& |\Delta^{\prime}|^2(\cos^2\theta 
+ \frac{M^2}{\mu_{\rm eff}^2}\sin^2{\theta}).
\label{effective}
\eea

The antiparticles states $E^+$ are obtained replacing $\mu$ by $-\mu$ and 
the coupling constants are taken as $G_1=3/4G_1$ and $G_2=3/16G_2$, from instanton induced interactions. 
  
\begin{figure}
\centering
\includegraphics[width=0.8\columnwidth,angle=-90]{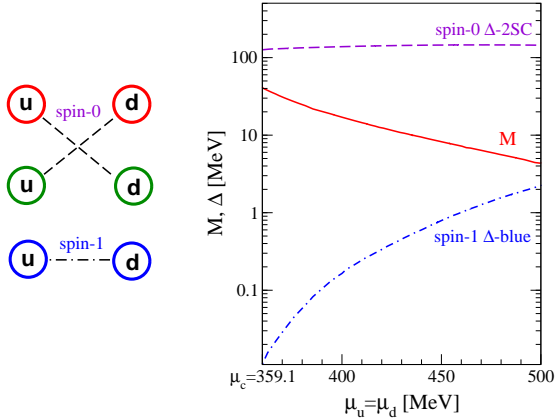}
\caption{Spin-0 2SC + spin-1 of the {\it blue} quarks pairing pattern (left panel)  and the corresponding dynamical mass and energy gaps (right panel) for symmetric two-flavor quark matter, $\mu_u=\mu_d$. 
Parameters: $\Lambda=595.5$ MeV, $G\Lambda^2=3.37$, $m=5.56$ MeV. Fixed $M=380$ MeV. 
}
\label{fig:1}      
\end{figure}

In Fig. \ref{fig:1} we show the results of solving the gap equations (\ref{gap_equations})
 for $M_f$, 
$\Delta$ and $\Delta^{\prime}$.   
%%%%%%%%%%%%%%%%%%%%%%%%%%%%%%%%%%%%%%%%%%%%%%%%%%%%%%%%%%%%%%%%%%%%%%%%%%%%%%%%%%%%
%%%%%%%%%%%%%%%%%%%%%%%%%%%%%%%%%%%%%%%%%%%%%%%%%%%%%%%%%%%%%%%%%%%%%%%%%%%%%%%%%%%%
The gaps $\Delta^{\prime}$   are strongly $\mu$-dependent rising functions 
and typically of the order of magnitude of keV, at least two orders of magnitude 
smaller than the corresponding 2SC gaps. For the nonlocal extension, the gaps are even smaller, not larger than 
$\Delta^{\prime} \approx 0.05$ MeV in the $\mu$-range shown in Fig.~\ref{fig:1}. 
Such small gaps will have no influence on the equation of state and in this pattern 
the are no quarks that remain unpaired. 
Nevertheless, it is unlikely that the blue quark pairing could survive the 
constraint imposed by charge neutrality: the Fermi seas 
of the up and down quarks should differ by about 50-100 MeV and this is much 
larger than the magnitude of the gap in the symmetric case.

 \subsection{{\it Likely:} Color Spin Locking phase in compact stars}
\label{sec_likely_2SC}
Single flavor spin-1 pairs are good candidates since they are inert 
against large splittings in the quark Fermi levels for different 
flavors caused by charge neutrality.
They  have been introduced first in   
\cite{Schafer:2000tw} and \cite{Alford:2002rz} and their properties have been  
investigated later more in detail, see  \cite{Schmitt:2004hg} and more recently \cite{Alford:2005yy}. 
Among many possible pairing patterns (polar, planar, A, CSL), the transverse Color-Spin-Locking  phase (CSL) 
has been demonstrated to be the ground state of a spin-1 color superconductor at $T=0$ having the largest pressure (\cite{Schmitt:2004et}). Their non-relativistic limit reproduces the B-phase in $^3$He, which locks angular momentum and spin and is also the most stable phase at $T=0$. We will show the features that make the CSL likely to occur in the interior of compact stars. 

\paragraph{Single flavor and color neutral condensates.}

\begin{figure}[bth]
\centering
\includegraphics[width=0.75\columnwidth,angle=-90]{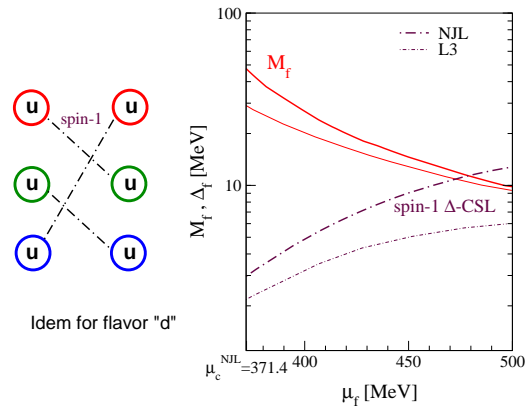}
\caption{Spin-1 CSL pairing (left panel) and the corresponding dynamical mass and energy gaps (right panel as a function of a single flavor chemical potential $\mu_f$, with $f=u,d$). For the NJL model (thick lines) the same parameterization as Fig.~\ref{fig:1} is used. For the nonlocal extension (label L3), parameters are fixed to $M=380$ MeV, see \cite{Aguilera:2006cj}.
}
\label{fig:2}       
\end{figure}

In the CSL condensates the color and spin are locked 
\be 
\langle q_f^{T}~C\gamma^3\lambda_2~q_f \rangle  
= \langle q_f^{T}~C\gamma^1\lambda_7~q_f \rangle  
= \langle q_f^{T}~C\gamma^2\lambda_5~q_f \rangle \equiv \eta_f
\label{csl}
\ee 
in an antisymmetric antitriplet in the color-space and an axial vector in the spin-space. A scheme of the pairing is shown in Fig~\ref{fig:2}, on the left. 
%%%%%%%%%%%%%%%%%%%%%%%%%%%%%%%%%%%%%%%%%%%%%%%%%%%%%%%%%%%%%%%%%%%%%%%%%%%%%%%%%%%%
%%%%%%%%%%%%%%%%%%%%%%%%%%%%%%%%%%%%%%%%%%%%%%%%%%%%%%%%%%%%%%%%%%%%%%%%%%%%%%%%%%%%
One can see immediately, that since Cooper pairs in the CSL phase are single flavor the results will not be affected by charge neutrality and thus we have overcome one of the most restrictively constraints of quark matter in compact stars. Moreover, there are equal number of color antitriplets ($\bar r$, $\bar g$, $\bar b$) and color neutrality is automatically fulfilled. The CSL diquark gaps  are 
\be 
\Delta_f = 4 H_v \eta_f~, 
\label{Delta_f} 
\ee 
where $H_v=3/8G$ from Fierz-transformation of single gluon exchange. 

One finds that the different flavors ($u,d$) decouple (\cite{Aguilera:2005tg}) and the
thermodynamic potential is given by the sum
$\Omega_q(T,\mu) =  \Omega_u(T,\mu_u) +  \Omega_d(T,\mu_d) $ where 
\bea  
\Omega_f(T,\mu_f)&=&  
\frac{1}{8G} (M_f-m)^2
+ \frac{3}{8H_v} |\Delta_f|^2
\nonumber\\ 
&-&
\sum_{k=1}^6 \int \frac{d^3p}{(2\pi)^3}  
\left(E_{f;k}+2T\ln{(1+e^{-E_{f;k}/T})}\right).~
%\nonumber\\ 
\label{omega} 
\eea  

\paragraph{Absence of ungapped modes}

Defining the effective variables as modified by the factor $M_f/\mu_f$ that counts for finite mass effects
\bea 
\varepsilon_{f, {\rm eff}}^2 &=& \vec p\,^2+M_{f,{\rm eff}}^2~,\\
M_{f,{\rm eff}} &=& \frac{\mu_f}{\mu_{f,{\rm eff}}}M_f~, \\
\mu_{f,{\rm eff}}^2 &=& \mu_f^2+|\Delta_f|^2~,\\
\Delta_{f,\rm eff} & = &\frac{M_f}{\mu_{f,{\rm eff}}}\,|\Delta_f|~ 
\label{Deltaeff}
\eea 
we can express the dispersion relation  $E_1$ for the first particle mode  in the standard form 
\bea  
E_{f;{1}}^2&=& (\varepsilon_{f,{\rm eff}}-\mu_{f,{\rm eff}})^2+\Delta_{f,{\rm eff}}^2~,
\eea
while for the other particle modes the following approximation can be used
\bea  
 E_{f;3,5}^2&\simeq& (\varepsilon_{f}-\mu_{f})^2+  c_{f;3,5}^{(1)}\,|\Delta_{f}|^2 .
\label{E_1,2} 
\eea
The dispersion relations for the antiparticles ($k=2,4,6$) can be obtained  replacing $\mu$ by $-\mu$ and the corresponding coefficients $c_{f;3,5}^{(1)}$ by $c_{f;4,6}^{(1)}$ (see \cite{Aguilera:2005tg} for a complete treatment). 

\begin{figure}
\centering
\includegraphics[width=0.75\columnwidth,angle=-90]{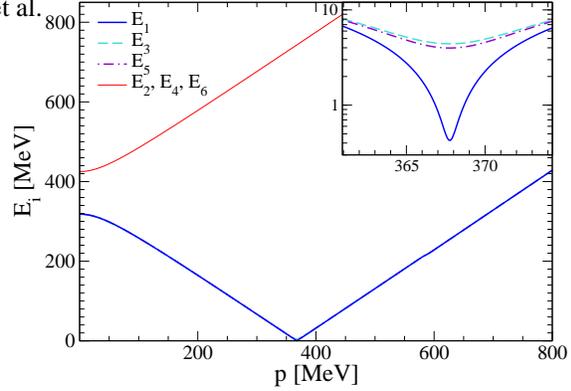}
\caption{Excitation energies in the CSL phase as a function
    of the momentum at  $\mu_c=371.4$ MeV, corresponding to the Fig.~\ref{fig:2}. 
Particle modes $E_1, E_3, E_5$ are zoomed in the upper corner on the right 
showing that there are no gapless modes. 
Antiparticle modes $E_2, E_4, E_6$ are superimposed and shown as thin lines. 
NJL model 
}
\label{fig:3}       
\end{figure}

The results of solving the gap equations (\ref{gap_equations}) for the CSL phase are shown in Fig.~\ref{fig:2} (right panel). The dynamical quark mass and the diquark 
gap are plotted as functions of $\mu_f$. Their 
corresponding excitation energies $E_i$, $i=1-6$ for at the critical chemical potential $\mu_c$, the onset for the CSL phase,  are shown in Fig.~\ref{fig:3}.
%%%%%%%%%%%%%%%%%%%%%%%%%%%%%%%%%%%%%%%%%%%%%%%%%%%%%%%%%%%%%%%%%%%%%%%%%%%%%%%%%%%%
%%%%%%%%%%%%%%%%%%%%%%%%%%%%%%%%%%%%%%%%%%%%%%%%%%%%%%%%%%%%%%%%%%%%%%%%%%%%%%%%%%%%%

From Fig.~\ref{fig:2} we see that the CSL gaps are strongly $\mu_f$-dependent  
functions 
in the considered domain.  Since the constituent mass in vacuum $M$ determines the 
$\mu_c$ at which the chiral phase transition takes place (and thus the onset for the superconducting phase), 
the low density region is qualitatively  
determined by the parameterization. This is crucial for the later matching of the quark sector 
with a high density nuclear matter EoS and for the construction of stable configurations of hybrid stars. 
Models with an onset of the quark matter phase at high densities might not give stable quark cores (\cite{Buballa:2003et}).  

 From Fig.~\ref{fig:3} we have learned that there are no gapless modes as 
a direct consequence of keeping the finite size of the mass of the $u,d$ quarks in 
(\ref{Deltaeff}).
This may play a crucial role suppressing the neutrino emissivities and preventing the direct Urca process to work. With unpaired quark species, the direct Urca is  so efficient that cool down the star too fast, in disagreement with observational data\footnote{although data might be also consistent with models that allow direct Urca in quark matter, lowering the core temperature but retain the surface temperature high enough due to e.g. Joule heating in the crust (\cite{Ponsinprogress}).}.

\subsection{Neutral CSL and critical temperatures}
\begin{figure}
\centering
\includegraphics[width=0.75\columnwidth,angle=-90]{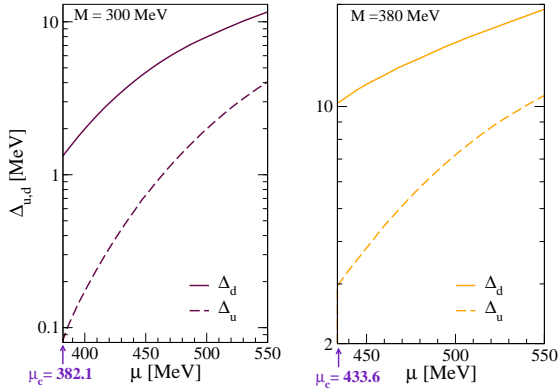}
\caption{Neutral matter: $\Delta_d$, $\Delta_u$ as a function of $\mu$. Two different parameterizations are considered: $M=300$ MeV on the left and $380$ MeV on the right. Note the different scales for $\Delta$  in both cases showing that $\Delta$ is very sensitivity to the variation of the constituent mass $M$ and the corresponding parameterization.}
\label{fig:4_0}       
\end{figure}

It is straightforward now to consider charge neutral matter introducing an electron chemical potential $\mu_e$, 
with $\mu_u=\mu-\frac{2}{3}\mu_e$  and $\mu_d=\mu+\frac{1}{3}\mu_e$, 
such that the total charge vanishes. Therefore, the effect of introducing charge neutrality is that the energy gap for the two flavors splits in two branches:  $\Delta_d$, $\Delta_u$, where $\Delta_u < \Delta_d$ for a given $\mu$ as it shown in Fig.~\ref{fig:4_0}.   
%%%%%%%%%%%%%%%%%%%%%%%%%%%%%%%%%%%%%%%%%%%%%%%%%%%%%%%%%%%%%%%%%%%%%%%%%%%%%%%%%%%%
%%%%%%%%%%%%%%%%%%%%%%%%%%%
The difference between them could be as large as a factor $\approx 10$ for low densities.  The onset for neutral CSL matter is displayed in both parameterizations considered.

\begin{figure}
\centering
\includegraphics[width=0.75\columnwidth,angle=-90]{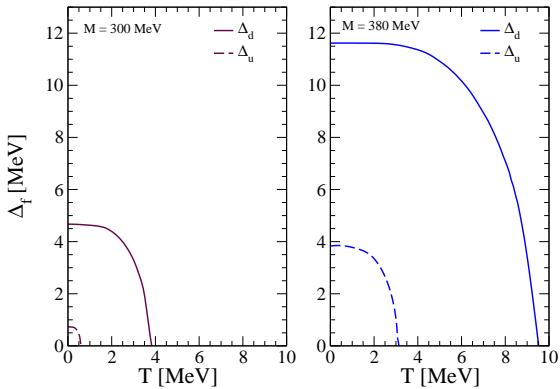}
\caption{Neutral matter: Temperature dependence of the CSL gaps. 
}
\label{fig:4}       
\end{figure}

\begin{figure}
\centering
\includegraphics[width=0.75\columnwidth,angle=-90]{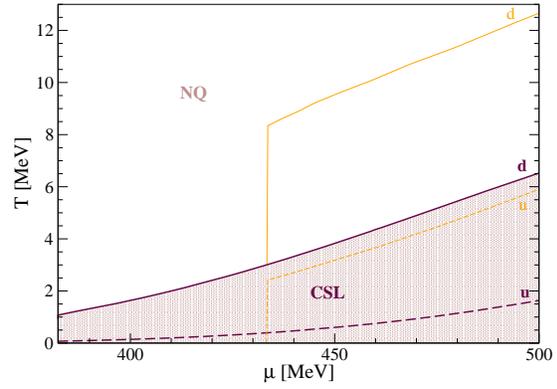}
\caption{Two flavor neutral matter at intermediate densities: critical temperatures are shown for the $d$ (solid line) and for $u$ (dashed line) quarks as a function of $\mu$.  
The thick lines correspond to the parameterization of NJL with 
$M=300$ MeV and the thin lines to $M=380$ MeV. 
}
\label{fig:5}       
\end{figure}

We analyze the temperature dependence of the gaps in Fig.~\ref{fig:4}. They show a decreasing behavior that ends at the critical temperature $T_c$ where a second order phase transition to normal (unpaired) quark matter occurs.  
%%%%%%%%%%%%%%%%%%%%%%%%%%%%%%%%%%%%%%%%%%%%%%%%%%%%%%%%%%%%%%%%%%%%%%%%%%%%%%%%%%%%
%%%%%%%%%%%%%%%%%%%%%%%%%%%%%%%%%%%%%%%%%%%%%%%%%%%%%%%%%%%%%%%%%%%%%%%%%%%%%%%%%%%%%
It is worth to notice that the critical temperature $T_c$ for the CSL phase  deviates from the BCS relation 
($T_c \approx 0.57\, \Delta\, (T=0)$),  as it has been demonstrated for  weak coupling in \cite{Schmitt:2004et}, and follows 
\be
{T_c}_f \approx 0.82\, \Delta_f\, (T=0)~. 
\ee
Our calculations confirm this result and are shown in the phase diagram of Fig.~\ref{fig:5}. 
%%%%%%%%%%%%%%%%%%%%%%%%%%%%%%%%%%%%%%%%%%%%%%%%%%%%%%%%%%%%%%%%%%%%%%%%%%%%%%%%%%%%
%%%%%%%%%%%%%%%%%%%%%%%%%%%
After the critical potential $\mu_c$ for the occurrence of CSL phase, ($\mu_c=382.1$ MeV for the $M=380$ MeV parameterization),  two lines labeling $T_c$ for the two flavor species appear: as quark matter cool down, first the $d$ quarks (solid line) will condense  and then, at lower temperatures, the phase transition for the $u$  quarks (dashed line) will take place. Thus, when the temperature falls below $\simeq 1$ MeV is likely that the two flavor quark matter will be in the CSL phase. 

Although a comparison of the CSL free energy  with the energy for other possible pairings remains to be done\footnote{cristallyne phases, phases with consistent treatment of the strange quark mass, etc}, and will decide whether the CSL phase is preferable, there are many indications that point in the same direction. 
First, NJL model calculations show that when a moderate coupling constant for the usual 2SC phase is considered, 
two flavor normal quark matter dominates the intermediate region of the phase diagram (\cite{Ruster:2005jc,Blaschke:2005uj}).  Then, recently in \cite{Alford:2005yy}, single pairing NJL model calculations performed with massless $u-d$ quarks and the mass of the $s$ quark taken through an effective chemical potential, show that for large strange quark mass, CSL dominates the phase diagram at low temperature and a second order phase transition to unpaired quark matter occurs as the temperature increases.  
So, we expect that when the densities in the interior of a compact star are high enough to allow for two-flavor quark matter but not so high to have the strange quark deconfined, 
and the temperature has fallen below $T_c$, a CSL quark core might develop. Moreover, for such small gaps the pressure is expected to be approximately the one of the normal $u-d$ quark matter and it has been obtained that hybrid stars with a relatively large normal quark matter core can be stable (\cite{Grigorian:2003vi}).

\subsection{Neutral CSL and magnetic field}

Let's suppose that in the core of a neutron star the density is high enough that a 
quark core could be formed and has already condensed to the  CSL phase. 
The final question we try to address is the nature of the interaction 
between the magnetic field present in a neutron star and the CSL quark core. 
Although this is a question that requires a careful analysis, the aim of this section is to review the work done on this matter and present estimations derivate from the  CSL gaps calculated before. 

For the interaction of the CSL phase with an external magnetic field $B$ it has been stated that  
the CSL phase exhibits an electromagnetic Meissner effect (\cite{Schmitt:2003xq}) since all the gluons and the photon acquires a non-vanishing Meissner mass proportional to the quark chemical potential (multiplied by the appropriate gauge coupling) 
(\cite{Schmitt:2003aa}). The formation of the Cooper pairs 
  breaks the symmetry (color-spin) $SU(3)_c \times SO(3)_J$ to $SO(3)_{c+J}$, that is a global symmetry.  
Concerning electromagnetism, it was shown that 
there is no symmetry subgroup left of  $SU(3)_c \times U(1)_{em}$ different from the trivial and thus 
there is no combination of color and electric charge for which the Cooper pairs are invariant

This is in contrast to the spin-0 2SC and CFL phases that although the condensate has non-zero electric charge, there is a residual local symmetry  $\tilde U(1)$ with an associated  lineal combination of the photon and a gluon (``rotated photon'') that remains massless.  The ``rotated'' $\tilde B$ field can penetrate and propagate in the 2SC and in the CFL phase (\cite{Alford:1999pb}).  Therefore, 2SC and CFL phases are not superconductors from the electromagnetic point of view and the magnetic field $\tilde B$ will penetrate without the restriction to be 
quantized to flux tubes and is stable over very long time scales.
On the other hand, $B$ cannot penetrate in the CSL phase unless it exceeds the critical magnetic field $B_{c}$ that destroy the superconducting phase. 

Similarly to the ordinary superconductors, we can estimate the penetration length as the inverse of the Meissner mass  
%of the corresponding ``mixture'' 
of the photon ($\gamma$) and/or the gluon ($a$) that would penetrate the matter following (\cite{Schmitt:2003aa})
\be
\lambda_{\gamma,a}\simeq \frac{1}{m_{\gamma,a}} \simeq \frac{1}{(e,g)\mu}
%=\frac{7.1} {10^{4} {\rm MeV}} =0.14 {\rm fm} 
\ee
where $e,g$ are their corresponding coupling constants in dense matter. At densities in neutron stars, photons are weakly coupled $e^2/4\pi \simeq 1/137$ 
while gluons are strongly coupled  $g^2/4\pi \approx 1$ and taking 
a typical value of $\mu \simeq 400$ MeV we obtain\footnote{Heaviside-Lorentz units are used: $\mu_0=\epsilon_0=1$, $\hbar=c=k=1, e=\sqrt{4\pi/137}\approx 0.3$, $g\approx 3.5$ and thus GeV$^2 = 5.10^{19}$ G} that the penetration length is
$\lambda \simeq 1-10$ fm. 

The coherence length is proportional to the inverse of the energy gap 
$\xi \simeq 1/\Delta$ and
from our previous results we obtained that 
being $\Delta \simeq 1$ MeV $\approx 1/200\, {\rm fm}$.  Then, the 
ratio 
\be 
\frac{\lambda}{\xi} \simeq \frac{1-10}{200} <<  1/\sqrt{2},  
\ee
confirms that CSL is a  Type-I superconductor and the magnetic field would be expelled from macroscopic regions if it does not exceed the critical field to destroy the condensate. This may be consistent with recent investigations that state that surface temperature anisotropies inferred from the observations are due to the crust confination of the magnetic field 
(\cite{Geppert:2004sd,Perez-Azorin:2005ds}). Moreover, studies of the light curves of neutron stars have shown evidence of precession and if this is confirmed, this might be inconsistent with Type II superfluity.  Because of the strong interaction between the rotational vortices and the flux tubes in the latter, Type I superconductivity may be required (\cite{Link:2003hq}). 

\begin{figure}
\centering
\includegraphics[width=0.75\columnwidth,angle=-90]{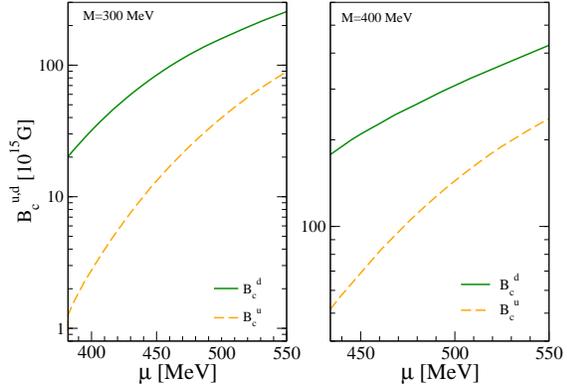}
\caption{Estimated critical $B$ as a function of $\mu$ for CSL neutral matter. Note the different scales for the two parameterizations used. 
}
\label{fig:6}       
\end{figure}

We can estimate the critical field $B_c$ using the Ginzburg-Landau approach following \cite{Bailin:1983bm}, 
\be
B_{c}^2=8\mu  p_F\frac{(k_BTt)^2}{7\zeta(3)}.
\ee
Assuming that due to the low mass of $u,d$,  the Fermi momentum $p_F \simeq \mu$ and if the $T << T_c$ then the parameter $t=(T-T_c)/T_c \simeq 1$. We can also express  $T_c \approx 0.82\, \Delta\, (T=0)$
%\footnote{$T_c$ for the CSL phase deviates from the BCS relation, $T_c \approx 0.57\, \Delta\, (T=0)$ see \cite{Schmitt:2004et}, and more recently \cite{Alford:2005yy}}, 
%$\zeta(3)\approx 1.2$, 
and consider the $\mu$-dependence of the gaps to obtain 
$B_{c}(\mu) \simeq 0.8\, \mu\, \Delta(\mu)$ and in relevant units 
\be
B_{c}(\mu) = 1.6 \,\left(\frac{\mu}{400 \rm MeV}\right) \,
\left(\frac{\Delta(\mu)}{\rm MeV}\right) \,10^{16} G~.
\ee
Therefore
for typical values of $\mu=400$ MeV and CSL gaps not exceeding $\Delta=1$  MeV, we have that $B_c \simeq 10^{16} G$.  
In Fig.~\ref{fig:6} we display $B_c(\mu)$ taking into account the density dependence of our previous  NJL model results.  
%%%%%%%%%%%%%%%%%%%%%%%%%%%%%%%%%%%%%%%%%%%%%%%%%%%%%%%%%%%%%%%%%%%%%%%%%%%%%%%%%%%%
%%%%%%%%%%%%%%%%%%%%%%%%%%%
As the figures show, $B_c$ is highly density dependent, and due to the different $\mu_u,\mu_d$  there is a splitting in the critical field for the two flavors: $B_c^u \leq B_c^d$. 

What are the consequences for compact stars?
First, due to the strong density dependence, we expect that if a CSL quark core forms, it begins to grow from the center of the star. Since it is a Type I superconductor, 
the magnetic field will be expelled from it. If $B$ is not so large that the CSL phase persist against the magnetic field ($B\leq B_c^u$), then, the characteristic times for the expulsion of the field over macroscopic regions and for the quark core grow start to compete. 
According to \cite{Ouyed:2003ge}, the magnetic field expulsion time over a sizable region ($\simeq 10$ m) can be evaluated as proportional to the electrical conductivity in the normal phase $\sigma_{el}\simeq 10^{23-24} s^{-1}$ (\cite{Shovkovy:2002sg}),
\be
\tau_{\rm exp}\simeq \left(\frac{\sigma_{\rm el}}{10^{23} {\rm s}^{-1}} \right)
\left( \frac{\delta}{10 {\rm m}}\right)^2 
\left( \frac{B}{10^{15} {\rm G}}\right)\,s
\ee 
getting typical values of seconds. An estimation of how fast the quark core develops is unknown, but if it is much smaller than $\tau_{\rm exp}$ it might not have time to expel $B$. 
Since CSL and the magnetic field do not coexist, $B$ would be frozen in an mixed state composed of alternating regions of normal material with flux density $H_c$ and superconducting material exhibiting Meissner screening. This would be also the case for intermediate fields, comparable with $B_c$.  For the case $B_u^c \leq B\leq B_d^c$, although $B$ will try to destroy the $u-$condensates,  $d$-currents will cancel the field inside, restoring the superconducting phase.  As a results, we speculate that the field structure will be rearranged to give the mixed state described above.  

The spin down estimates from  highly magnetized neutron stars ({\it magnetars}) show that the magnetic field could be as large as $B \simeq 10^{15}$ G (\cite{Woods:2004kb}). If one consider that they could  reach even higher values (one order of magnitude more?) as a results of flux conservation into confined regions, the scenario of a CSL quark core under the stress of magnetic fields that are of the order of $B_c$ seems not to be unrealistic.

%%%%%%%%%%%%%%%%%%%%%%%%%%%%%%%%%%%%%%%55
\section{Summary and outlook}
We summarize here the interesting features that the  CSL phase presents for compact star applications:
\begin{itemize}
\item since Cooper pairs are single flavor they are not  affected by charge neutrality 
\item color neutrality is automatically fulfilled 
\item there are no gapless modes. This has important consequences e.g. in neutrino emissivities, suppressing the direct Urca process that leads to a rapid cooling 
\item their small gaps ($\Delta \simeq 1$ MeV) will not influence on the EoS, and the pressure is expected to be approximately the one of the normal $ud$-quark matter without pairing. Similar to the latter, stable hybrid star configurations could be obtained with a relatively large quark matter core.  
\item it exhibits a Meissner effect and the magnetic field of a neutron star would be  expelled from a CSL quark core if it does not exceed the critical magnetic field $B_{c}$ that destroy the superconducting phase. This hypothesis might be consistent with recent investigations that indicate the crust confination of the magnetic field. Moreover, the hypothesis of a Type I superconductor is supported 
by the inferred precession of some compact objects.   
\end{itemize}

This study presents a qualitative analysis of the spin-1 pairing 
and cannot be taken as conclusive to decide whether this phase could be realized in neutron star cores. However,  the features listed show that CSL could be consistent with the phenomenology of compact stars in many aspects.

\begin{acknowledgements}
D.N.A thanks  J.~Pons and S.~Reddy for their interest in this work and for fruitful discussions on previous version of this manuscript. CSL model calculations has been done in collaboration with D.~Blaschke, M.~Buballa, N.~Scoccola and H.~Grigorian. D.N.A. thanks also D.~Page for interesting discussions during the conference.
\end{acknowledgements}

% BibTeX users please use
%\bibliographystyle{spmpsci}


\end{document}